\documentclass[12pt,a4wide]{article}
\usepackage{afterpage}
\usepackage{epsfig}
\usepackage[a4paper]{geometry}
\usepackage{float}
\usepackage[stable]{footmisc}
\usepackage[utf8x]{inputenc}
\usepackage{a4wide}
\usepackage{placeins}
\usepackage{amsmath,amssymb,amstext,amsthm,amssymb}
\usepackage{amsfonts}
\usepackage{mathrsfs}
\usepackage{framed}
\usepackage{graphicx}
\usepackage{bbold}
\usepackage{seqsplit}
\usepackage{slashed} 
\usepackage{tensor} 
\usepackage{braket} 
\usepackage{array}
\usepackage{graphicx}
\usepackage{enumerate}
\usepackage{setspace}
\usepackage{hyperref}
\usepackage{overpic}
\usepackage{bbm}
\usepackage{cite}
\usepackage{color}
\usepackage{lipsum}
\usepackage{bm}
\usepackage{tikz}

\DeclareSymbolFontAlphabet{\mathbb}{AMSb}

\let\baraccent=\=
\renewcommand{\=}[1]{\stackrel{#1}{=}}

\begin{document}

\pagestyle{plain}

\makeatletter
\@addtoreset{equation}{section}
\makeatother
\renewcommand{\theequation}{\thesection.\arabic{equation}}
\pagestyle{empty}
{\hfill \small MIT-CTP/5447}
\vspace{0.5cm}

\begin{center}
{\Large \bf{On D3-brane Superpotential}}\\
\vskip 9pt

\end{center}
\vspace{0.5cm}

\begin{center}
\scalebox{0.95}[0.95]{{\fontsize{14}{30}\selectfont Manki Kim$^{a}$}} \vspace{0.35cm}
\end{center}

\begin{center}
\vspace{0.25 cm}
\textsl{$^{a}$Center for Theoretical Physics, Massachusetts Institute of Technology, Cambridge, MA 02139, USA}\\

	 \vspace{1cm}
	\normalsize{\bf Abstract} \\[8mm]
\end{center}
\begin{center}
	\begin{minipage}[h]{15.0cm}
We study the D3-brane moduli dependence of the non-perturbative superpotential generated by the D7-brane gaugino condensation in a type IIB compactification on an orientifold of the elliptic Calabi-Yau threefold $\Bbb{P}_{[1,1,1,6,9]}[18].$ Building on the work of Ganor, in the weak coupling limit, we compute the D3-brane dependent one-loop pfaffian in the flat coordinates and find that the one-loop pfaffian is written as the Jacobi theta function. We also comment on the complex structure dependence of the D3-brane dependent one-loop pfaffian.
	\end{minipage}
\end{center}
\newpage
\setcounter{page}{1}
\pagestyle{plain}
\renewcommand{\thefootnote}{\arabic{footnote}}
\setcounter{footnote}{0}
%
%
\setcounter{tocdepth}{2}
\tableofcontents
\newpage
\section{Introduction}
Spacetime filling D3-branes are ubiquitous in the 4d $\mathcal{N}=1$ vacua of type IIB string compactifications. The prevalence of the D3-branes, therefore, necessitates understanding their roles in string compactifications. In particular, D3-brane moduli play pivotal roles in moduli stabilization \cite{Giddings:2001yu,Kachru:2003aw,Balasubramanian:2005zx} and model building \cite{Kachru:2003sx,Dasgupta:2002ew,Ruehle:2017one} through the mixing with the Kahler moduli \cite{DeWolfe:2002nn,Giddings:2005ff}.\footnote{Through the heterotic/F-theory duality, spacetime filling D3-branes are identified with vertical NS5-branes in heterotic string compactifications on elliptic Calabi-Yau manifolds \cite{Friedman:1997yq}. Therefore, understanding the dynamics of D3-branes may shed light on the non-perturbative physics of heterotic string theory.}

The first step to precisely characterize the physics of D3-branes is to compute the D3-brane moduli dependence of the Kahler potential and the superpotential. In the large volume limit and the weak string coupling limit, computation of the D3-brane Kahler potential amounts to the computation of the Kahler metric of the compactification manifold, for which a popular choice is an orientifold of a Calabi-Yau threefold. Computation of the Calabi-Yau metric is an active field of research \cite{donaldson2005some,Headrick:2005ch,Ashmore:2019wzb,Ashmore:2021ohf,Douglas:2020hpv,Anderson:2020hux,Larfors:2021pbb,Larfors:2022nep}. We will add nothing to this difficult computation of the Kahler potential in this draft. It is known that in supersymmetric compactifications the D3-brane moduli are perturbatively flat \cite{Giddings:2001yu} and are lifted by non-perturbative superpotentials generated by Euclidean D3-branes and D7-brane gaugino condensations \cite{Ganor:1996pe}. The goal of this draft is to further the understanding of the D3-brane moduli dependence of the non-perturbative superpotential.

The D3-brane moduli dependence of the non-perturbative superpotential arises from the one-loop pfaffian \cite{Ganor:1996pe}. Although the one-loop diagram between a spacetime filling D3-brane and the source of the non-perturbative superpotential is computable in principle \cite{Blumenhagen:2006xt,Alexandrov:2022mmy}, it is very difficult to do so in practice in a Calabi-Yau orientifold compactification because one needs the spectrum of the open string excitations. This is why the explicit computation of the D3-brane moduli dependence was only carried out either in toroidal compactifications \cite{Berg:2004ek} or in non-compact compactifications \cite{Baumann:2006th}, because only in such cases were the explicit metrics of the compactification manifolds are accessible. 

But, not knowing the explicit Calabi-Yau metric shall not stop us from attempting to compute the D3-brane moduli dependence of the non-perturbative superpotential. With the help of holomorphy \cite{Seiberg:1994bp}, one can still compute the exact superpotential which may be perturbatively inaccessible. For example, in the context of heterotic string theory, the vector bundle moduli dependence of the worldsheet instanton superpotential \cite{Witten:1999eg} was computed by carefully examining the analytic structure of the one-loop pfaffian \cite{Buchbinder:2002pr}.\footnote{This computation was conjectured to be extended to the seven-brane moduli dependence of the non-perturbative superpotential in F-theory \cite{Cvetic:2012ts}.}

In this spirit, building on the pioneering work by Ganor \cite{Ganor:1996pe}, by analyzing its analytic structures we compute the D3-brane moduli dependence of the non-perturbative superpotential generated by the D7-brane gaugino condensation in a type IIB compactification on an orientifold of the elliptic Calabi-Yau threefold $\Bbb{P}_{[1,1,1,6,9]}[18]$ \cite{Candelas:1994hw}.

Before we conclude the introduction, we highlight a puzzle that arises from the result of \cite{Ganor:1996pe}. By the work of Ganor \cite{Ganor:1996pe}, it is understood that if a non-perturbative superpotential term is generated by a Euclidean M5-brane wrapping on a smooth rigid divisor defined as a vanishing locus of the unique section of a line bundle $Z=0,$ the one-loop pfaffian dependence on a mobile M2-brane is determined to be
\begin{equation}
\mathcal{A}_{M2}=Z|_{M2}\,,
\end{equation}
where $Z|_{M2}$ is the section evaluated at the M2-brane locus. This result is confusing. In the combination of the F-theory limit \cite{Vafa:1996xn}, and the weakly coupled type IIB limit \cite{Sen:1996vd,Sen:1997gv}, the one-loop pfaffian is computed by the one-loop diagram between a mobile D3-brane and the source of the non-perturbative superpotential. The spectrum of the open string excitations depends on complex structure moduli, and hence one finds that in explicit computations the D3-brane dependent one-loop pfaffian is sensitive to the complex structure moduli of the compactification manifold \cite{Berg:2004ek}. Because it is not obvious how the section $Z$ encodes the complex structure dependence of the D3-brane dependent one-loop pfaffian, we find a tension between the two results.

In this draft, we resolve this puzzle. The main idea comes from the observation that in \cite{Berg:2004ek} the explicit computation of the one-loop pfaffian was performed using the flat coordinates, not the section. We show that, with the help of the Weierstrass p-function of the elliptic fibration, the section $Z$ can be related to the flat coordinates. The main result is \eqref{eqn:pfaffian res}. And as a result, we find the perfect agreement between the result of \cite{Ganor:1996pe} and \cite{Berg:2004ek}.

The organization of this paper is as follows. In \S\ref{sec:example}, we construct an O3/O7 orientifold of the Calabi-Yau threefold $\Bbb{P}_{[1,1,1,6,9]}[18].$ We show that the non-perturbative superpotential is generated by the gaugino condensation on a non-Higgsable SO(8) seven-brane stack. We argue that in the absence of mobile D3-branes, the non-perturbative superpotential term in this model is independent of the complex structure moduli. In \S\ref{sec:ganor}, we compute the D3-brane dependent one-loop pfaffian building on the work of \cite{Ganor:1996pe}. We express the one-loop pfaffian in the flat coordinates and find that the one-loop pfaffian is a Jacobi theta function in the weak string coupling limit. In \S\ref{sec:orbifold limit}, we take an orbifold limit to match the result we find in \S\ref{sec:ganor} to the result of \cite{Berg:2004ek} and find the perfect agreement. In \S\ref{sec:conclusion}, we conclude.
\section{An example}\label{sec:example}
In this section, we construct an O3/O7 orientifold of $X_3:=\Bbb{P}_{[1,1,1,6,9]}[18].$ $V_4:=\Bbb{P}_{[1,1,1,6.9]}$ is a $\Bbb{P}_{[2,3,1]}$ fibration over $\Bbb{P}^2,$ and therefore $X_3$ is an elliptic fibration over $\Bbb{P}^2.$ The detailed study of this Calabi-Yau in the context of mirror symmetry was carried out in \cite{Candelas:1994hw}. A GLSM of $X_3$ is given by
\begin{equation}
\left(
\begin{array}{cccccc}
y_1&y_2&y_3&X&Y&Z\\
1&1&1&6&9&0\\
0&0&0&2&3&1
\end{array}
\right)\,,
\end{equation}
where $y_i$ are homogeneous coordinates of the base manifold $\Bbb{P}^2,$ and $X,~Y,~Z$ are homogeneous coordinates of $\Bbb{P}_{[2,3,1]}.$ The Stanley-Reisner ideal of this toric variety is
\begin{equation}\label{eqn:SR ideal}
SR(V_4)=\{y_1y_2y_3,XYZ\}\,.
\end{equation}
The only toric divisor with no normal deformation is $[Z=0].$ 

We define the Calabi-Yau threefold $X_3$ as the vanishing locus of 
\begin{equation}
Y^2=4X^3-g_2 XZ^4-g_3Z^6\,,
\end{equation}
where $g_2$ and $g_3$ are polynomials of base degree $12$ and $18,$ respectively. It is useful to note that $Z=Y=0$ does not intersect this Calabi-Yau manifold. The discriminant of the elliptic fibration is 
\begin{equation}
\Delta:=g_2^3-27g_3^2\,,
\end{equation}
and the j-invariant is given by
\begin{equation}
j:=1728 \frac{g_2^3}{\Delta}\,.
\end{equation}
The j-invariant enjoys small $q=e^{2\pi i\tau}$ expansion
\begin{equation}
j=q^{-1}+744+196884q+21493760q^2+\dots\,.
\end{equation}

With the Weierstrass form, we can study the flat coordinates on the elliptic fiber. Given a point $[y_1,y_2,y_3]$ on the base manifold $\Bbb{P}^2,$ the flat coordinate $z$ of the elliptic fiber is related to $[X,Y,Z]$ via
\begin{equation}
\wp(z,\tau(y))=\frac{X}{Z^2}\,,
\end{equation}
and
\begin{equation}
\wp'(z,\tau(y))=\frac{Y}{Z^3}\,,
\end{equation}
where $\wp(z,\tau(y))$ is the Weierstrass elliptic p-function
\begin{equation}
\wp(z,\tau(y)):=\frac{1}{z^2}+\sum_{(n,m)\in \Bbb{Z}^2\backslash(0,0)}\left(\frac{1}{(z-n-m\tau(y))^2}-\frac{1}{(n+m\tau(y))^2}\right)\,.
\end{equation}
The inverse map of the Weierstrass elliptic p-function enjoys nice properties
\begin{equation}\label{eqn:inverse wp1}
\wp^{-1}(Y=0)=\left\{ z=\frac{1}{2}\right\}\cup\left\{ z=\frac{\tau}{2}\right\}\cup\left\{z=\frac{1+\tau}{2}\right\}\,,
\end{equation}
and
\begin{equation}\label{eqn:inverse wp2}
\wp^{-1}(Z=0)=\{z=0\}\,.
\end{equation}
\eqref{eqn:inverse wp1} implies that $[Y=0]\in X_3$ is a three-point fibration over $\Bbb{P}^2,$ and similarly $[Z=0]\in X_3$ is topologically equivalent to $\Bbb{P}^2.$ In the next section, the Weierstrass p-function will be used heavily.

To construct an orientifold, we consider a $\Bbb{Z}_2$ involution $\mathcal{I}:Y\mapsto -Y.$ Note that due to the toric equivalence relation
\begin{equation}
[X,Y,Z]\sim [\lambda^2 X,\lambda^3 Y,\lambda Z]\,,
\end{equation}
for $\lambda\neq0,$ the $\Bbb{Z}_2$ involution $\mathcal{I}$ has an equivalent but a different presentation $\mathcal{I}:Z\mapsto -Z.$ We find that no geometric modulus is projected out by the orientifolding
\begin{equation}
h^{1,1}_+=h^{1,1}=2,~h^{2,1}_-=h^{2,1}=272\,.
\end{equation}
The orientifold $\mathcal{B}_3:=X_3/\mathcal{I}$ has two O7-plane loci
\begin{equation}
O7=\{Y=0\}\cup \{Z=0\}\,.
\end{equation} 
One may be tempted to conclude that there is an O3-plane at $y_1=y_2=y_3=0,$ as one can relate the $\Bbb{Z}_2$ involution $Y\mapsto -Y$ to $[y_1,y_2,y_3]\mapsto - [y_1,y_2,y_3]$ by using the toric equivalence relation
\begin{equation}
[y_1,y_2,y_3,X,Y]\sim [\lambda y_1,\lambda y_2,\lambda y_3,\lambda^6 X,\lambda^9Y]\,,
\end{equation}
for $\lambda\neq0.$ But, $y_1=y_2=y_3=0$ is in the SR-ideal \eqref{eqn:SR ideal} hence does not have a solution. As a result, we conclude that there is no O3-plane in $\mathcal{B}_3.$

\begin{figure}
\centering
  \begin{tikzpicture}[scale=.7]
    \draw (-2,0) node[anchor=east]  {$D_4$};
	\draw (-1.1,-1.55) node[anchor=east]  {$1$};
    \draw (1.85,-1.55) node[anchor=east]  {$1$};
    \draw (0.4,-0.7) node[anchor=east]  {$2$};
    \draw (-1.1,0.2) node[anchor=east]  {$1$};
    \draw (1.85,0.2) node[anchor=east]  {$1$};
    \draw[xshift=0 cm,thick] (0 cm,0) circle (.3cm);
    \draw[xshift=0 cm,thick] (30: 17 mm) circle (.3cm);
    \draw[xshift=0 cm,thick] (-30: 17 mm) circle (.3cm);
    \draw[xshift=0 cm,thick] (30:- 17 mm) circle (.3cm);
    \draw[xshift=0 cm,thick] (-30: -17 mm) circle (.3cm);
    \draw[xshift=0 cm,thick] (30: 3 mm) -- (30: 14 mm);
    \draw[xshift=0 cm,thick] (-30: 3 mm) -- (-30: 14 mm);
    \draw[xshift=0 cm,thick] (30: -3 mm) -- (30: -14 mm);
    \draw[xshift=0 cm,thick] (-30: -3 mm) -- (-30:- 14 mm);
  \end{tikzpicture}
\caption{The affine Dynkin diagram for $D_4.$ }\label{fig:affine D4}
\end{figure}
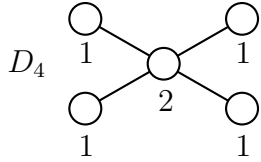

We first analyze gauge groups of the D7-brane stacks in the weakly coupled type IIB description. We will work in the up-stair picture, meaning the tadpole cancellation requires $[D7]=8[O7].$ The D7-brane is wrapped on the vanishing locus of the following polynomial
\begin{equation}
\Delta_{D7}:=Y^2Z^2 B_6-B_4^2\,,
\end{equation}
where $B_6$ is a polynomial of bi-degree $(54,24)$ and $B_4$ is a polynomial of bi-degree $(36,16).$ It appears that $B_6$ can be written as $B_6= Z^6 \tilde{B}_6,$ and similarly $B_4$ can be written as $B_4=Z^4\tilde{B}_4.$ As a result, we write
\begin{equation}
\Delta_{D7}= Z^8(Y^2\tilde{B}_6-\tilde{B}_4^2)\,.
\end{equation}
As a result, at a generic point in the moduli space, the elliptic fiber has $D_4$ singularity on $Z=0.$ The associated monodromy cover for the $D_4$ singularity on $Z=0$ is \cite{Grassi:2011hq}
\begin{equation}
\psi^3+\psi\left(-\frac{1}{3}Y^4+2\tilde{B}_4|_{Z=0}t\right)+\left(\frac{2}{27} Y^6-\frac{2}{3}Y^2\tilde{B}_4|_{Z=0}t+\tilde{B}_6|_{Z=0}t^2\right)=0\,,
\end{equation}
where $t$ is a small parameter describing the axio-dilaton $t=\mathcal{O}(e^{\pi i \tau}).$ We find that there is only one monomial in $\tilde{B}_4$ that survives the restriction onto $Z=0$
\begin{equation}
\tilde{B}_4|_{Z=0}=\alpha Y^4\,.
\end{equation}
Similarly, we find 
\begin{equation}
\tilde{B}_6|_{Z=0}=\beta Y^6\,.
\end{equation}
Because $Z=Y=0$ does not have a solution, we can safely fix $Y=1$ by toric rescaling relation. As a result, the monodromy cover is
\begin{equation}\label{eqn:monodromy cover}
\psi^3+\psi\left(-\frac{1}{3}+2\alpha t\right)+\left(\frac{2}{27} -\frac{2}{3}\alpha t+\beta t^2\right)=0\,.
\end{equation}
There are three solutions of \eqref{eqn:monodromy cover}. Hence, we find that the gauge group on $Z=0$ is $SO(8)$ at a generic D7-brane configuration. As $Z=0$ is topologically $\Bbb{P}^2$ there is no adjoint matter realized on this non-Higgsable SO(8) gauge theory.\footnote{For detailed studies of the non-Higgsable clusters, see \cite{Morrison:2012js,Morrison:2014lca,Halverson:2015jua}.} Furthermore, at a generic point in the moduli space, there is no gauge enhancement locus on $Z=0.$ Therefore we find that this non-Higgsable SO(8) gauge theory is pure SYM.

To study the gaugino condensation of the non-Higgsable SO(8) gauge theory at $Z=0,$ it is useful to consider M-theory and F-theory descriptions. To do so, we construct an elliptic Calabi-Yau fourfold $Y_4$ as the Weierstrass model over the orientifold $\mathcal{B}_3.$ One can think of $\mathcal{B}_3$ as a $\Bbb{P}^1$ fibration over $\Bbb{P}^2$ with the twisting $L=6[H]$ where $[H]$ is a hyperplane class in $\Bbb{P}^2.$ A GLSM of this Calabi-Yau fourfold is by
\begin{equation}
\begin{array}{c|c|c|c|c|c|c|c}
y_1&y_2&y_3&X&\tilde{Z}&\widehat{X}&\widehat{Y}&\widehat{Z}\\\hline
1&1&1&6&0&18&27&0\\\hline
0&0&0&1&1&4&6&0\\\hline
0&0&0&0&0&2&3&1
\end{array}
\end{equation}
where $\tilde{Z}$ is identified with $Z^2.$ This Weierstrass model $Y_4$ is singular, because of the non-Higgsable SO(8) stack at $\tilde{Z}=0.$ 

Let us first study M-theory compactification on $Y_4.$ In 3d $\mathcal{N}=2$ theories, pure Super-Yang-Mills theories contain Coulomb branches unlike 4d $\mathcal{N}=1$ theories. Geometrically, moving away from the singularity in the moduli space along the Coulomb branch corresponds to blowing up along the singularity of the elliptic fiber. In the Coulomb branch, we can understand the gaugino condensation via Euclidean M5-brane instantons \cite{Katz:1996fh,Katz:1996th}.

Resolving the singularity along $S:=[Z]$ introduces exceptional $\Bbb{P}^1$'s in the fiber which we denote by $e_i$ for $i=1,\dots,4.$ We denote the affine $\Bbb{P}^1$ in the fiber direction by $e_0.$ Each $\Bbb{P}^1$ corresponds to a Dynkin node in the corresponding affine Dynkin diagram. Let $a_i$ be the Dynkin label for the Dynkin node $e_i,$ then we have
\begin{equation}
\sum_i a_i e_i=[\Bbb{E}]\,,
\end{equation}
where we have $\sum_ia_i=C_2(so(8))=6.$ For the Dynkin diagram and the Dynkin labels of $D_4,$ see Fig.\ref{fig:affine D4}.  Because $S$ is smooth, it is rather straightforward to check if the superpotential by a Euclidean M5-brane wrapped on $D_i =S\times e_i$ following the analysis of \cite{Witten:1996bn}.\footnote{For the recent progress in understanding the superpotential generated by Eulcidean M5-branes wrapped on singular divisors, see \cite{Gendler:2022qof}.} Assuming that there is no M2-brane on the divisor $S,$ a sufficient condition for the generation of the superpotential is
\begin{equation}\label{eqn:suff cond}
h^\bullet(D_i,\mathcal{O}_{D_i})=(1,0,0,0)\,.
\end{equation}
Because $S$ is $\Bbb{P}^2,$ every $D_i$ has a topology of $\Bbb{P}^2\times \Bbb{P}^1$ whose hodge vector is 
\begin{equation}
h^\bullet(\Bbb{P}^2\times \Bbb{P}^1,\mathcal{O}_{\Bbb{P}^2\times \Bbb{P}^1})=(1,0,0,0)\,,
\end{equation}
Hence a Euclidean M5-brane wrapped on $D_i$ satisfies the sufficient condition \eqref{eqn:suff cond} to generate a term in the superpotential
\begin{equation}\label{eqn:np EM5}
W_{np}\supset\sum_i\mathcal{A}_{D_i}e^{-\frac{1}{l_M^6}V_{S} V_{e_i}+i\int_{D_i} C_6}\,,
\end{equation}
where $l_M$ is the M-theory unit length, $V_S$ and $V_{e_i}$ are volume of $S$ and $e_i$ respectively.

In general, the one-loop pfaffian $\mathcal{A}_{D}$ can depend on all but Kahler moduli. But, a simplification can occur. As was explained in \cite{Witten:1996hc,Ganor:1996pe}, in the \emph{absence} of spacetime filling M2/D3-branes, the one-loop pfaffian $\mathcal{A}_{D}$ is a section of a line bundle $\mathcal{L}$ of the intermediate Jacobian $\mathcal{J}(D):=H^3(D,\Bbb{R})/H^3(D,\Bbb{Z}),$ where the Chern-class of $\mathcal{L}$ is the principal polarization of $\mathcal{J}(D).$ As for a rigid divisor $D,$ $h^{3,0}(D)$ is zero, we have
\begin{equation}
\dim\mathcal{J}(D)=2h^{2,1}(D)\,.
\end{equation}
Hence, if a rigid divisor $D$ also satisfies $h^{2,1}(D)=0,$ $\mathcal{A}_D$ is pure constant in the \emph{absence} of mobile M2/D3-branes. When $\dim H^3(D)=0,$ we call $D$ a pure divisor.\footnote{This condition was heavily used in a recent study of moduli stabilization \cite{Demirtas:2021nlu}.} Because $h^{2,1}(\Bbb{P}^2\times\Bbb{P}^1)=0,$ this essentially shows that the superpotential term \eqref{eqn:np EM5} is independent of complex structure moduli of $Y_4,$ in the absence of mobile M2/D3-branes.\footnote{Note also that one can verify this claim using the toric technique \cite{Kim:2021lwo}.}

So far we have studied the superpotential in the Coulomb branch, which does not exist in F-theory compactification. So, even though the conclusion on the superpotential we drew from M-theory compactification is highly suggestive that the superpotential term should be generated in the F-theory description, we shall check this claim.

To recover the superpotential in the F-theory limit, we follow the beautiful argument illustrated in \cite{Denef:2008wq}. Taking the F-theory limit is the small fiber volume limit $V_{\Bbb{E}}\rightarrow 0$ while keeping 
\begin{equation}
\frac{l_s^4}{l_M^6}V_{\Bbb{E}}
\end{equation} 
fixed, where $l_s$ is the unit string length. So, to take the F-theory limit for the superpotential \eqref{eqn:np EM5}, we need to relate $V_{e_i}$ to volume of the elliptic fiber. This can be achieved by minimizing the superpotential \eqref{eqn:np EM5} subject to the constraints
\begin{equation}
\sum_i a_i V_{e_i}=V_{\Bbb{E}}\,,~ \sum_{i}\int_{D_i} C_6=\int_{S\times\Bbb{E}}C_6\,.
\end{equation}
After solving the Lagrange multiplier equations, one finds
\begin{equation}
W_{np}=6\,\mathcal{A} \,e^{-\frac{1}{6}\left(\frac{1}{l_M^6}V_SV_{\Bbb{E}}+i\int_{S\times \Bbb{E}}C_6+2\pi m\right)}\,,
\end{equation}
where we defined $\mathcal{A}:=2^{-1/3}\prod_i \mathcal{A}_i^{a_i/6} $ and $m\in \Bbb{Z}.$ Taking the F-theory limit, we recover the gaugino condensation superpotential
\begin{equation}\label{eqn:gaugino condensation}
W_{np}=6\,\mathcal{A}\,e^{-\frac{1}{6 }\left(\frac{1}{l_s^4}V_S+i\int_S C_4+2\pi m\right)}\,.
\end{equation}
As we showed that $\mathcal{A}_i$ for all $i$ is independent of the complex structure moduli of $Y_4$ in the absence of mobile M2/D3-branes, we conclude that the gaugino condensation superpotential \eqref{eqn:gaugino condensation} is also independent of the complex structure moduli in the absence of mobile M2/D3-branes.

Until now, we have been ignoring the consistency constraints: the tadpole cancellation condition \cite{Sethi:1996es} 
\begin{equation}\label{eqn:tadpole}
N_{D3}+\frac{1}{2}\int G_4\wedge G_4=\frac{\chi(\tilde{Y}_4)}{24}\,,
\end{equation}
the flux quantization condition \cite{Witten:1996md,Freed:1999vc}
\begin{equation}\label{eqn:quantization}
G_4+\frac{1}{2}c_2(\tilde{Y}_4)\in H^{2,2}(\tilde{Y}_4)\cap H^4(\tilde{Y}_4,\Bbb{Z})\,,
\end{equation}
and the compatibility of the fourform flux with the Poincare symmetry \cite{Dasgupta:1999ss}. Because $\chi(\tilde{Y}_4)\neq0,$ there cannot be a consistent vacuum with no fourform flux and no mobile D3-branes. In the next section, to make the orientifold compactification picture precise we will take the weak coupling limit. Furthermore, we will assume that there is a consistent choice of the fourform flux and the number of mobile D3-branes that satisfy \eqref{eqn:tadpole} and \eqref{eqn:quantization}, and study the implications of the mobile D3-branes on the one-loop pfaffian of the non-perturbative superpotential.\footnote{In the global Sen limit \cite{Halverson:2017vde,Kim:2022jvv}, an algorithm to find a consistent choice of the flux was demonstrated in \cite{Demirtas:2021nlu}. }

\newpage
\section{Ganor prefactor as a section of a line bundle}\label{sec:ganor}
In the previous section, we studied the gaugino condensation superpotential generated by the non-Higgsable SO(8) stack at $Z=0$ and concluded that the corresponding one-loop pfaffian is independent of the complex structure moduli of $Y_4$ in the \emph{absence} of mobile M2/D3-branes.

As the reader may have noticed, the pure rigidity of $D$ does not imply that $\mathcal{A}_D$ is a pure constant if there are mobile M2/D3-branes. In fact, it is by now very well known that $\mathcal{A}_D$ depends on position moduli of mobile M2/D3-branes \cite{Ganor:1996pe,Berg:2004ek,Baumann:2006th}. Let us consider a mobile D3-brane, which is separated from the divisor $S$ by a properly measured distance $z.$ The lowest mode of the string extended between the mobile D3-brane and $S$ is a state with mass $z$ in the fundamental representation of the D7-brane gauge group on $S.$ Using the standard analysis based on the holomorphy of the superpotential \cite{Intriligator:1995au}, one arrives at 
\begin{equation}\label{eqn:D3-superpotential}
\mathcal{A}_{D3}\simeq z^{1/6}\,,
\end{equation}
around $z\simeq0$. 

Because \eqref{eqn:D3-superpotential} was obtained in the limit where $z\simeq0,$ the next step we shall take is to generalize this formula for an arbitrary value of $z.$ As was explained in \cite{Ganor:1996pe}, the local structure of \eqref{eqn:D3-superpotential} dictates that the sixth power of the one-loop pfaffian $\mathcal{A}_D^6$ is a section of the line bundle $[Z].$ Hence, the form of the pfaffian should be 
\begin{equation}
\mathcal{A}_{D3}^6=Z_{D3}\,,
\end{equation}
where $Z_{D3}$ is the section $Z$ evaluated at the D3-brane position. Quite interestingly, $\mathcal{A}_{D3}$ is not a section of an honest line bundle. Clearly, this one-loop pfaffian shows that if the mobile D3-brane is on top of the D7-brane stack at $Z=0,$ the non-perturbative superpotential vanishes due to the new zero mode.

This form of the pfaffian, albeit unique and well justified, raises a puzzle. Because the one-loop pfaffian dependence on the D3-brane moduli comes from the mass of excited states of the string stretched between the D3-brane and the D7-brane, one expects that the one-loop pfaffian to depend on complex structure moduli of the Calabi-Yau manifold as well. This expectation was confirmed by explicit computations \cite{Berg:2004ek,Kim:2018vgz}. But, it is not obvious how the section $Z_{D3}$ of the line bundle $[Z]$ encodes the complex structure dependence.

To understand this obvious puzzle, we would like to take one possibility very seriously. In all the known explicit computations that show that the one-loop pfaffian $\mathcal{A}_{D3}$ depends both on the D3-brane moduli and the complex structure moduli of the Calabi-Yau, it was the case that the D3-brane moduli were expressed in terms of the flat coordinates. For the explicit computations of the one-loop pfaffian, it is of course more convenient to use the flat coordinates, but the usage of the flat coordinates can obscure the nature of the line bundles. Furthermore, the relation between the flat coordinate and the homogeneous coordinates, or the Cox ring, can be quite complicated, and in general, depends on the complex structure moduli of the Calabi-Yau manifold. So, one naive guess could be that although the pfaffian is just proportional to the section $Z,$ when expressed in terms of the flat coordinates, one sees the dependence on complex structure moduli.

Now, to test this proposal, we must perform a few consistency checks. Of these, the zeroth-order check is to confirm that $Z$ when written in terms of the flat coordinates is proportional to $\vartheta_{1}$ in the orbifold limit.  

First, we recall the fundamental theorem that states that every elliptic function has finitely many poles in the fundamental domain and the sum of its residue is zero. Hence, if we are to relate the flat coordinates to $Z,$ we must find a quasi-periodic modular function at best. We also note that in the convention where the discriminant $\Delta$ is a modular form of weight 12, the homogeneous coordinates $X,~Y,~Z$ are expected to be modular forms of  weights 2, 3, 0, respectively. As a result, the quasi-periodic function in question should have a weight 0.

Although we have now understood the modular properties of $Z,$ given that there can be many modular forms with zero weight, it is not yet easy to pin down the answer. To help the search, it is informative to recall the Weierstrass form and the Weierstrass p-function
\begin{equation}
Y^2=4X^3-g_2XZ^4-g_3Z^6\,,
\end{equation}
\begin{equation}\label{eqn:WP series}
\wp(z,\tau(y)):=\frac{1}{z^2}+\sum_{(n,m)\in \Bbb{Z}^2\backslash(0,0)}\left(\frac{1}{(z-n-m\tau(y))^2}-\frac{1}{(n+m\tau(y))^2}\right)\,,
\end{equation}
where $z$ is the flat coordinate of the elliptic fiber. The Weierstrass p-function parametrizes the elliptic fiber via the flat coordinates, as it satisfies a differential equation
\begin{equation}
(\partial_z\wp(z,\tau(y)))^2=4\wp(z,\tau(y))^3-g_2 \wp(z,\tau(y))-g_3\,.
\end{equation}
This implies that when $Z\neq0,$ one can relate $Y$ and $X$ to the flat coordinate via
\begin{equation}
\frac{Y}{Z^3}=\partial_z\wp(z,\tau(y))\,,
\end{equation}
\begin{equation}
\frac{X}{Z^2}=\wp(z,\tau(y))\,.
\end{equation}
This claim that the Weierstrass p-function parametrizes the elliptic curve may come as a surprise, as from \eqref{eqn:WP series} it does not look quite obvious that the Weierstrass p-function should have the weight 2. In fact, there exists a more informative form of the Weierstrass p-function that will illustrate many properties we would like to use \cite{olver2010nist}
\begin{equation}\label{eqn:WP theta}
\wp(z,\tau(y))=\left(\frac{\pi \vartheta_2(0,q)\vartheta_3(0,q)\vartheta_4( z,q)}{\vartheta_1( z,q)}\right)^2-\frac{\pi^2}{3}\left(\vartheta_2^4(0,q)+\vartheta_3^4(0,q)\right)\,.
\end{equation}
From \eqref{eqn:WP theta}, one can directly see that the Weierstrass p-function has weight 2, confirming our expectation. We can take a derivative of the Weierstrass p-function,
\begin{align}
\partial_z\wp(z,\tau(y))=&-2\left(\pi \vartheta_2(0,q)\vartheta_3(0,q)\right)^2\left(\frac{\vartheta_4( z,q)}{\vartheta_1 ( z,q)}\right)^3 \frac{\vartheta_4(0,q)^2\vartheta_2( z,q)\vartheta_3( z,q)}{\vartheta_4^2( z,q)}\\
=&-2 (\pi \vartheta_2(0,q)\vartheta_3(0,q)\vartheta_4(0,q))^2\frac{\vartheta_2( z,q)\vartheta_3( z,q)\vartheta_4( z,q)}{\vartheta_1( z,q)^3}\,.\label{eqn:WP derivative}
\end{align}
In addition to the information that $\partial_z\wp(z,\tau(y))$ is a modular function of weight 3, \eqref{eqn:WP derivative} contains a few more useful lessons to teach us. The zero of the flat coordinate $z=0$ is in one to one correspondence with the vanishing locus $Z=0.$ Hence, the orientifold action in the flat coordinate $\mathcal{I}_z: z\mapsto -z$ should be equivalent to the orientifold action $\mathcal{I}:Z\mapsto -Z.$ In fact, for this equivalence to hold, $Y=0$ should correspond to three points $z=\frac{1}{2},~\frac{\tau}{2},~\frac{1+\tau}{2}.$ As the simple zero of $\vartheta_1( z,q)$ corresponds to $Z=0,$ poles of $\partial_z\wp(z,\tau(y))$ cannot be interpreted as the zeros of $Y=0.$ Rather, zeros of $\vartheta_i( z,q)$ for $i=2,~3,~4$ should correspond to the zeros of $Y=0.$ Quite nicely, zeros of $\vartheta_i(\pi z,q)$ for $i=2,~3,~4$ are $z=\frac{1}{2},~\frac{\tau}{2},~\frac{1+\tau}{2}.$ 

Now, given the relations \eqref{eqn:WP theta} and \eqref{eqn:WP derivative}, we proceed to find the expression of $Z$ in the flat coordinate $z.$ As we have studied already, $Z$ should be of the form
\begin{equation}
Z= f(q) \vartheta_1(z,q)\,,
\end{equation}
where $f(q)$ is a modular form of weight $-1/2.$ The choice of $f(q)$ is a gauge choice one can freely choose from, provided that the constraints \eqref{eqn:WP theta} and \eqref{eqn:WP derivative} are met. Note that this gauge transformation corresponds to the Kahler transformation. We therefore without loss of generality choose a gauge in which 
\begin{equation}
f(q)= \left(2^{-1} \vartheta_1(0,q)\vartheta_2(0,q)\vartheta_3(0,q)\right)^{-1/3}=\eta(q)^{-1}\,.
\end{equation}
The reason for choosing this gauge will be more manifest in the next section. In this gauge, we have
\begin{equation}
X=\frac{\vartheta_1(z,q)^2}{\eta(q)^2}\wp(z,\tau(y))\,,~Y=-8\pi^2\eta(q)^3\vartheta_2(z,q)\vartheta_3(z,q)\vartheta_4(z,q)
\end{equation}
and
\begin{equation}
Z=\frac{\vartheta_1(z,q)}{\eta(q)}\,.
\end{equation}
This determines the one-loop pfaffian uniquely up to the gauge transformation
\begin{equation}\label{eqn:pfaffian res}
\mathcal{A}_{D3}^6=\frac{\vartheta_1(z,q)}{\eta(q)}\,.
\end{equation}

By finding the expression \eqref{eqn:pfaffian res}, we have now resolved the puzzle we posed. Expressed as a section of a line bundle, the one-loop pfaffian need not explicitly depend on complex structure moduli of the Calabi-Yau manifold. But, when one expresses the section in terms of the flat coordinates, as was computed in explicit examples, the one-loop pfaffian dependence on the Calabi-Yau complex structure moduli becomes explicit.

This result \eqref{eqn:pfaffian res} is in a way quite surprising. It is worth stressing that to arrive at \eqref{eqn:pfaffian res}, the only information we took into the derivation was essentially the structure of the sections of the line bundles. As this derivation does not rely on any metric information of the Calabi-Yau manifold, it is expected to hold even for a generic fibration. Thus, a deviation from the answer one can obtain from a toroidal orbifold model could've been expected. But, unlike what one may expect, the result we found \eqref{eqn:pfaffian res} is structurally completely the same as the one-loop pfaffian obtained in the orbifold model. We will explicitly check this claim in the next section.

\newpage
\section{Orbifold limit}\label{sec:orbifold limit}
In the previous section, we found a general expression of the one-loop pfaffian dependence on D3-brane moduli in the elliptic fibration over $\Bbb{P}^2$ by carefully examining the structure of the line bundles. In this section, we take a limit in complex structure moduli in which $X_3$ becomes $(\text{K3}\times \text{T}^2)/\Bbb{Z}_2$ and directly compute the one-loop pfaffian dependence to show that the proposed formula passes a non-trivial check. 

We take a slightly different convention for the Weierstrass form in this section
\begin{equation}
Y^2=X^3+f XZ^4+gZ^6\,,
\end{equation}
where $f$ and $g$ are polynomials of degree 12 and 18, repectively. In this convention, the discriminant and the j-invariant are
\begin{equation}
\Delta=4f^3+27g^2\,,
\end{equation}
\begin{equation}
j=1728 \frac{4f^3}{\Delta}\,.
\end{equation}
To take an orbifold limit, one can take an ansatz for $f$ and $g$ such that the j-invariant of the elliptic fibration is a constant. We therefore take an ansatz
\begin{equation}
f= -\frac{1}{3} h^2\,,
\end{equation}
\begin{equation}
g=\left(\frac{2}{27}-t\right) h^3\,,
\end{equation}
where $h$ is a degree 6 polynomial in the base coordinates. We find that the discriminant and the j-invariant evaluate to
\begin{equation}
\Delta=t(-4+27t)h^6\,,
\end{equation}
\begin{equation}
j=-\frac{256}{t(-4+27t)}\,.
\end{equation}
Analogous to Sen's orientifold limit, we can introduce a degree 3 auxiliary coordinate $\xi$ to understand the base manifold is a two-point fibration over $\Bbb{P}^2,$ which is a K3-manifold
\begin{equation}
\xi^2=h\,.
\end{equation} 
Note that this K3-manifold can be understood as an anti-canonical hypersurface in $\Bbb{P}_{[1,1,1,3]}.$ Phrased differently, we find that in this orbifold limit, $X_3$ becomes $(\text{K3}\times \text{T}^2)/\Bbb{Z}_2.$

Same as before, we take the orientifold action $\mathcal{I}:Z\mapsto -Z.$ As a result, we find two O7-plane stacks $Y=0$ and $Z=0.$ One cautionary remark is in order. At a generic point in the moduli space, $Y=0$ is an irreducible divisor of the Calabi-Yau manifold. But, in the orbifold limit, $Y=0$ is a solution to the equation
\begin{equation}\label{eqn:Y=0 orbifold}
X^3-\frac{1}{3} h^2 XZ+\left(\frac{2}{27}-t\right) h^3 Z^6=0\,,
\end{equation}
which has three independent roots. The solutions to \eqref{eqn:Y=0 orbifold}, in the flat coordinates, as expected, are given as
\begin{equation}
z=\frac{1}{2},~\frac{\tau}{2},~\frac{1+\tau}{2}\,.
\end{equation}
As $\tau$ does not vary over the base manifold, there is no non-trivial monodromy to mix these three roots to form an irreducible divisor. Hence, $Y=0$ is three copies of the K3-manifold. 

Now, we proceed to compare the one-loop pfaffian we obtained \eqref{eqn:pfaffian res} with the explicit computation of the one-loop pfaffian performed in the literature \cite{Berg:2004ek,Baumann:2006th,Kim:2018vgz}. 

In general, computation of the one-loop pfaffian of the non-perturbative superpotential is very involved. But, it can be simplified in the case of gaugino condensation, as the one-loop pfaffian of the superpotential enjoys the interpretation of the log of the one-loop corrections to the holomorphic gauge coupling. Because the D7-brane stack in question fully wraps the K3-manifold, and therefore its low-energy gauge coupling is determined by the volume of the K3-manifold, to determine the one-loop pfaffian of the superpotential one can simply compute the one-loop correction to the K3 volume. 

For the one-loop pfaffian dependence on the D3-brane moduli, the relevant scale for the one-loop pfaffian is the mass of the string extended between the D3-brane and the D7-brane gaugino stack. Because the K3 manifold in the orbifold model is orthogonal to the tori, and the D7-brane wraps the entire K3, the geodesic length between a mobile D3-brane and the D7-brane can be computed without knowing the explicit K3 metric. This effectively allows one to read off the backreaction caused by a mobile D3-brane to the geometry.

As the one-loop pfaffian is the one-loop exact quantity that at the same time does not depend on Kahler moduli, to simplify the computation of the one-loop pfaffian we can take a weak string coupling limit and the large volume limit. Essentially in this limit, we can treat a mobile D3-brane as a point source in $\text{T}^2$ when the mobile D3-brane is seen from the D7-brane stack. Because the tree level holomorphic D7-brane gauge coupling is in an appropriate unit the K3 volume, all we have to evaluate is how the metric at the D7-brane locus in $\text{T}^2$ is perturbed by the point source away from the D7-brane stack. From this line of argument, one concludes that the one-loop pfaffian dependence on the D3-brane moduli, in the flat coordinates, shall take a form
\begin{equation}
\mathcal{A}_{D3}^6=\exp \left( 2\pi G(z;0)|_{\text{hol}} \right)\,,
\end{equation}
where $G(z;0)|_{\text{hol}}$ is the holomorphic part of the Green's function in $\text{T}^2$
\begin{equation}
G(z;0)|_{\text{hol}}=\frac{1}{2\pi}\log\left(\frac{\vartheta_1(z,q)}{\eta(q)}\right)\,.
\end{equation}
As a result, we obtain the one-loop pfaffian
\begin{equation}
\mathcal{A}_{D3}^6=\frac{\vartheta_1(z,q)}{\eta(q)}\,,
\end{equation}
which agrees with \eqref{eqn:pfaffian res}.

\section{Conclusions}\label{sec:conclusion}
In this work, building on the work of \cite{Ganor:1996pe}, we computed the D3-brane moduli dependence of the one-loop pfaffian of the non-perturbative superpotential in a type IIB compactification on an orientifold of the elliptic Calabi-Yau threefold $\Bbb{P}_{[1,1,1,6,9]}[18].$ With the help of the Weierstrass p-function, we expressed the D3-brane dependent prefactor in the flat coordinates at a generic point in the complex structure moduli and showed that the D3-brane dependent prefactor not only depends on the D3-brane moduli but also on the complex structure moduli of the compactification manifold. This result agrees with the explicit computations performed in \cite{Berg:2004ek,Baumann:2006th}. To compute the D3-brane dependent one-loop pfaffian, we heavily used the fact that the Calabi-Yau threefold $\Bbb{P}_{[1,1,1,6,9]}[18]$ is elliptic. It will be interesting to extend this work to more general Calabi-Yau manifolds. 

The other limitation of this work is that the result \eqref{eqn:pfaffian res} was obtained in the weak string coupling limit in which one can ignore the effects of the seven-brane moduli. Because the one-loop pfaffian is one-loop exact, and due to the fact that the spectrum of the open string excitations of the annuli diagrams between mobile D3-branes and the source of the non-perturbative superpotential does not depend on the D7-brane moduli, a naive expectation is that the D7-brane moduli will not enter the one-loop pfaffian. It will be worthwhile to explicitly compute the effect of D7-brane moduli away from the weak string coupling limit.

\section*{Acknowledgements}
We thank Jakob Moritz, Liam McAllister, Sergey Alexandrov for useful discussions and comments. We thank Yuji Tachikawa for inspirational questions. The work of MK was supported by Pappalardo Fellowship.

\bibliography{refs}
\bibliographystyle{JHEP}
\end{document}